\documentclass[pdflatex,sn-mathphys-num]{sn-jnl}

\usepackage{graphicx}%
\usepackage{multirow}%
\usepackage{amsmath,amssymb,amsfonts}%
\usepackage{amsthm}%
\usepackage{mathrsfs}%
\usepackage[title]{appendix}%
\usepackage{xcolor}%
\usepackage{textcomp}%
\usepackage{manyfoot}%
\usepackage{booktabs}%
\usepackage{algorithm}%
\usepackage{algorithmicx}%
\usepackage{algpseudocode}%
\usepackage{listings}%

\begin{document}

\title[Article Title]{Atemporality from Conservation Laws of Physics in Lorentzian-Euclidean Black Holes}

\author*[1]{\fnm{Silvia} \sur{De Bianchi}}\email{silvia.debianchi@unimi.it}

\author[2,3,4]{\fnm{Salvatore} \sur{Capozziello}}\email{capozziello@na.infn.it}

\author[5]{\fnm{Emmanuele} \sur{Battista}}\email{ebattista@lnf.infn.it}

\affil*[1]{\orgdiv{Department of Philosophy}, \orgname{University of Milan}, \orgaddress{\street{via Festa del Perdono 7}, \city{Milan}, \postcode{20122}, \country{Italy}}}

\affil[2] {\orgdiv{Dipartimento di Fisica ``Ettore Pancini'', Complesso Universitario di Monte S. Angelo, Universit\`a degli Studi di Napoli ``Federico II'', Via Cinthia Edificio 6, 80126 Napoli, Italy}}

\affil[3] {\orgdiv{Istituto Nazionale di Fisica Nucleare, Sezione di Napoli, Complesso Universitario di Monte S. Angelo, Via Cinthia Edificio 6, 80126 Napoli, Italy}}

\affil[4] {\orgdiv{Scuola Superiore Meridionale, Largo San Marcellino 10, 80138 Napoli, Italy}}

\affil[5]{\orgdiv{Istituto Nazionale di Fisica Nucleare, Laboratori Nazionali di Frascati, 00044 Frascati, Italy}}


\abstract{Recent results have shown that singularities can be avoided from the general relativistic standpoint in Lorentzian-Euclidean black holes by means of the transition from a Lorentzian to an Euclidean region where time loses its physical meaning and becomes imaginary. This dynamical mechanism, dubbed ``atemporality'', prevents the emergence of black hole singularities and the violation of conservation laws. In this paper, the notion of atemporality together with a detailed discussion of its implications is presented from a philosophical perspective. The main result consists in showing that atemporality is naturally related to conservation laws.}




\maketitle

\section{Introduction}\label{sec1}

In his beautiful masterpiece, {\it Bangs, Crunches, Whimpers, and Shrieks: Singularities and Acausalities in Relativistic Spacetimes}, John Earman spelled out the philosophical implications of Einstein's relativity \cite{Earman1995-EARBCW-3}. The  book focuses on the reconstruction of the problem of singularities that plagued general relativity (GR) and the strong connection of singularities with the notion of acausality. Earman's work influenced a generation of philosophers of physics and can be considered as a fundamental contribution for the understanding of the foundations of relativity. In the last decade, indeed, the physics of black holes, with regard to both relativistic and quantum effects attracted attention of philosophers (see \cite{curiel2019many} and references therein), giving rise to a lively debate on the foundations of black holes physics beyond the problem of singularities. However, the latter still plagues relativity, despite physicists’ attempts to remove them (see for instance \cite{ellis1992change}, \cite{ellis1992covariant}), resulting in a deeper reflection on the foundations of relativity and causality, as well as on the implications for both Quantum Cosmology and a theory of Quantum Gravity.\footnote{Indeed, one of the most relevant objectives of Quantum Gravity approaches, stemming for their distinct character from GR framework, is to show that they are successful in removing singularities, not only in black hole physics, but also in cosmological models (see \cite{li2023loop}, \cite{vidotto2022time}, \cite{oriti2023complex}, \cite{gielen2023stationary}).}  
Among the best-known procedures to avoid singularities, we find the {\it Wick rotation} \cite{Zee:2010qce}. The latter is widely employed in quantum mechanics and statistical mechanics to find out solutions in Minkowski spacetime from solutions in Euclidean space. It consists in a transformation substituting an imaginary variable, i. e., the time variable, with a real one. It is called ``{\it rotation}'' because  complex numbers can be represented in a plane as vectors and the multiplication of a complex number by the imaginary unit is equivalent to a rotating vector. 
In General Relativity, it implies the insertion of an imaginary time beyond the event horizon of a singularity and it was also applied in the context of the Hartle-Hawking No Boundary Conditions in Quantum Cosmology \cite{Hartle:2008ng}. In proposals appealing to the insertion of imaginary time through the {\it Wick rotation}, one finds models that mainly rely on inserting {\it ad hoc} mechanisms resulting in toy models that necessarily imply {\it ad hoc} solutions, thereby casting doubts on the effective possibility of signature change and singularity avoidance in GR and in some models of Quantum Cosmology. It is beyond our scope to deal in detail with approaches removing the singularity in Quantum Gravity and Quantum Cosmology, but with respect to this debate, we want to draw attention on \cite{Capozziello:2024ucm}, where a solution of the Einstein field equations represents a Lorentzian-Euclidean black hole.\footnote{Our main objective in this paper is to present the conceptual implication of the Lorentzian-Euclidean black hole and atemporality rather than showing whether other black hole spacetimes, e.g. Kerr BH are consistent.} In this self-gravitating system, the singularity is naturally avoided by means of a mechanism that is defined in terms of {\it atemporality}. This concept bears with it relevant consequences for theoretical physics and the philosophy of science, which we are going to spell out in this paper. The Lorentzian-Euclidean solution differs from Schwarzschild’s because in the former the singularity is avoided in a natural way: it is worth saying that the structure is geodesically complete and the dynamical mechanism labelled as `atemporality' allows one to define a signature-change transition from a Lorentzian to an Euclidean region.
Before proceeding, it is worth considering that the solution presented in \cite{Capozziello:2024ucm} is one of a broader class where singularities can be avoided. This class of solutions is fully consistent with the presence of conservation laws and constitutes the bulk of an approach that one can dub  ``{\it Singularity--Free Physics}''. 
The remarkable result presented in \cite{Capozziello:2024ucm} is that in the relativistic context, the singularity does not appear in the model of the black hole, because no classical in-falling massive body can reach the singularity. Any massive particle stops at the event horizon because, if entering it, conservation laws would be violated.
As we shall see, atemporality can be defined as the dynamical mechanism avoiding that an infalling body, approaching the event horizon, assumes an imaginary radial velocity. And it cannot be otherwise, if one does not want to violate causality and energy conservation in a self-gravitating system like a black hole. 
In this paper, we explore the fascinating idea that the Lorentzian-Euclidean black hole is only a particular case in which atemporality operates. The latter is a more general concept descending from conservation laws. It is worth noting that the concept of atemporality can find fruitful applications in other areas of philosophy of science even beyond the specific example of stationary black holes.
We shall proceed as follows: Section 2 is a brief summary of the Lorentzian-Euclidean black hole solution presented by \cite{Capozziello:2024ucm}. In Section 3, we spell out the concept of atemporality presented in \cite{Capozziello:2024ucm} and relate it with the concept of measurement.\footnote{Throughout the paper, we always refer to relativistic measurement and assume that time is dependent on reference frames.} Section 4 is devoted to the discussion of  conservation laws in order to show that atemporality descends directly from  Noether symmetries. In Section 5, after stating the general theorem of atemporality, we discuss the perspective of a {\it Singularity--Free Physics}. Concluding remarks are reported in Section 6.

\section{The Lorentzian-Euclidean Black Hole}
\label{sec2}

In \cite{Capozziello:2024ucm}, a solution of the vacuum Einstein field equations dubbed ``Lorentzian-Euclidean black hole'' points to the fact that the singularity can be avoided because the Kretschmann curvature invariant results to be finite and not exploding as for standard black holes.
This peculiarity is due to the fact that a massive particle approaching the event horizon cannot enter it otherwise its time becomes imaginary. Furthermore, a proper observer and a coordinate observer take an infinite (real) time to reach the horizon. Thus, \textbf{they always remain causally connected in the exterior solution which is geodesically complete}. 

It is worth noticing that the standard Schwarzschild metric is a particular case of this more general class of Lorentzian-Euclidean solutions showing the signature change: the metric assumes the usual Lorentzian signature outside the event horizon, it is degenerate at $r=2M$, and it displays an Euclidean signature at $r<2M$. Within this picture, which will be scrutinized in its foundations in another paper, the event horizon is understood as change surface, where the transition between the Lorentzian and Euclidean regimes occurs. 
According to the Schwarzschild coordinates $\{t,r,\theta,\phi\}$,  the Lorentzian-Euclidean Schwarzschild metric results as
\begin{align}
  {\rm d}  s^2 = - \varepsilon \left( 1-\frac{2M}{r} \right) {\rm d} t^2  
+ \dfrac{{\rm d} r^2 }{\left(1-\frac{2M}{r}\right)} 
+ r^2\Omega^2,
\label{Lorentzian-Euclidean-Schwarzschild}
\end{align}
where  $\Omega^2 =  {\rm d} \theta^2  + \sin^2 \theta \; 
{\rm d} \phi^2 $ and
\begin{align}
\varepsilon = {\rm sign} \left( 1 - \frac{2M}{r}\right)= 2 H \left( 1 - \frac{2M}{r}\right)-1, 
\label{epsilon-of-r}
\end{align}
and the step function $H\left(1-2M/r\right)$ is normalized in such a way that $H(0)=1/2$. The metric undergoes a signature change at the event horizon. Indeed, the function $\varepsilon$ allows one to divide the spacetime manifold $V$ into two regions $V_+$ and $V_-$ (i.e. $V = V_+ \cup V_-$) whose common boundary is represented by the change surface.

\begin{align}
\Sigma \; : r=2M,
\label{change-surface}
\end{align}
coinciding with the event horizon. The domain $V_+$ is characterized by the value $\varepsilon=1$ and pertains to the Lorentzian regime, where the metric is hyperbolic. At $\Sigma$, where $\varepsilon=0$, the metric becomes degenerate being
\begin{align}
{\rm det}\, g_{\mu \nu}=  -\varepsilon \left(r^2 \sin \theta \right)^2.
\label{determinant-metric}
\end{align}
In $V_-$, where $\varepsilon=-1$, the metric assumes an ultrahyperbolic signature and exhibits Euclidean structure, i.e. it has the same features as the Euclidean Schwarzschild metric.  
Technically, the Lorentzian-Euclidean Schwarzschild metric can be written in the Gullstrand-Painlev\'e coordinates \cite{Capozziello:2024ucm} and it does not blow up at the change surface \eqref{change-surface}. 
Thus, the change surface \eqref{change-surface} does not represent a surface layer and no impulsive wave is generated in spacetime. This implies that it is possible to study the behavior of orbits approaching the event horizon. In particular, radially in-falling massive particles, due to their simple structure, allow one to deduce some crucial aspects of the Lorentzian-Euclidean black hole geometry. 

For example, the Lorentzian-Euclidean metric implies a degeneration on the change surface \eqref{change-surface}. However, it is possible to regularize the Riemann tensor and obtain a well-behaved Einstein tensor, which amounts to zero. This means that the Lorentzian-Euclidean metric can be seen as a vacuum solution for an extended version of general relativity where degenerate metrics are considered. 
Thus, the dynamics of bodies radially approaching the event horizon can be investigated for both freely falling particles and accelerated observers. Furthermore, the coordinates through which we represent the system are those of gravitational waves, therefore the representational content and the representational machinery coincide in our work. What we describe is thus a signature change (transition from Lorentzian to Euclidean region) that is naturally obtained rather than manually inserted by means of a Wick rotation that generates imaginary time. 

The equations pertaining to the radial geodesic motion are
\begin{align}
\dot{r}^2 &= \varepsilon \left(\frac{2M}{r}\right) - \varepsilon \left(1-\frac{E^2}{\varepsilon^4}\right),
\label{radial-geod-1}
\\
\dot{t} &= \frac{E}{\varepsilon^2} \left( \frac{1}{1-2M/r}\right).
\label{radial-geod-2}
\end{align}
Let us consider an observer  starting at rest from some \emph{finite} distance $r_i>2M$. Then, from Eq. \eqref{radial-geod-1}, we obtain
\begin{align}
r_i = \frac{2M}{1-E^2},
\label{initial-radial-distance}
\end{align}
which represents a positive-definite length when
\begin{align}
0<E^2<1.    
\label{energy-constraint}
\end{align}
A first important outcome of the model follows immediately from Eq. \eqref{radial-geod-1} that $\dot{r}$ attains imaginary values as soon as $r<2M$, or equivalently $\varepsilon=-1$. We can further investigate this point following the standard analysis of \cite{Chandrasekhar1985}. Thus, the radial variable can be written by means of the relation
\begin{align}
r(\eta)= r_i \cos^2 \left(\eta/2\right),      
\end{align}
where $\eta \in [0,\eta_H]$, $\eta_H < \pi$ being the value \footnote{In standard general relativity, one usually assumes $\eta \in [0,\pi]$. However, as explained in this section,  radial geodesics cannot enter the black hole horizon.} of $\eta$ when $r=2M$. Therefore, the equations governing infalling radial geodesics take the form
\begin{align}
\dot{r} &= -\sqrt{\frac{\varepsilon^4 \sin^2 \left(\eta/2\right) + E^2 \left[\cos^2 \left(\eta/2\right) - \varepsilon^4\right] }{\varepsilon^3 \cos^2 \left(\eta/2\right)}},
\label{radial-geod-3}
\\
\dot{t}&= \frac{E}{\varepsilon^2} \frac{\cos^2 \left(\eta/2\right)}{\cos^2\left(\eta/2\right)-\left(1-E^2\right)}.
\label{radial-geod-4}
\end{align}
In view of the constraint \eqref{energy-constraint},  one observes  that the radial velocity  \eqref{radial-geod-3} assumes imaginary values when $\varepsilon=-1$. The same conclusion also applies to the following derivatives:
\begin{align}
\frac{\sigma}{\eta} &= \left(\dot{r}\right)^{-1} \frac{r}{\eta }= r_i \sin \left(\eta/2\right) \cos^2 \left(\eta/2\right) \sqrt{\frac{\varepsilon^3}{\varepsilon^4 \sin^2 \left(\eta/2\right) +E^2 \left[\cos^2 \left(\eta/2\right) - \varepsilon^4\right]}},
\label{d-sigma-d-eta}
\\
\frac{t}{\eta} &= \dot{t} \, \frac{\sigma}{ \eta} = \frac{E }{\varepsilon^2} \frac{r_i \cos^4 \left(\eta/2\right) \sin \left(\eta/2\right)}{\cos^2 \left(\eta/2\right) - \left(1-E^2\right)} \sqrt{\frac{\varepsilon^3}{\varepsilon^4 \sin^2 \left(\eta/2\right) +E^2 \left[\cos^2 \left(\eta/2\right) - \varepsilon^4\right]}},
\label{d-t-d-eta}
\end{align}
which become imaginary if $r<2M$. 

Let us now consider the motion of a particle that gets to the event horizon. When $\varepsilon=0$, one naively obtains, from Eqs. \eqref{radial-geod-1} and \eqref{radial-geod-2}, or equivalently Eqs. \eqref{radial-geod-3} and \eqref{radial-geod-4}, that both $\dot{r}$ and $\dot{t}$ diverge (note that also in the standard Lorentzian-signature pattern $\dot{t}$ blows up at the horizon). However, the behaviour of $\dot{r}$ is in contrast with the situation in which the motion starts at rest far away from the black hole. In this case,  $E=\varepsilon^2$ and the radial velocity \eqref{radial-geod-1} vanishes on the change surface and becomes imaginary inside the black hole. 

This issue can be solved if we recall that, in the model, the energy is defined as $E=-\varepsilon g_{\mu \nu} \xi^\mu u^\nu$, $\xi^\mu$ being the static Killing vector field. This means that for a given motion having  $2M<r_i<\infty$, one can write $E^2 =  \alpha^2 \varepsilon^4$, where $\alpha^2$ is some positive-definite bounded function depending on Eq. \eqref{energy-constraint}. Thus, either from Eq. \eqref{radial-geod-1} or Eq. \eqref{radial-geod-3}, one sees that $\dot{r}$ becomes zero on the change surface, just like in the scenario having $r_i \gg  2M$.

Therefore, infalling particles freely moving in the radial direction have a velocity $\dot{r}$ which vanishes at the event horizon and becomes imaginary after crossing it. This can be interpreted as an indication that the singularity at $r=0$ can be evaded because the observer never reaches it. This scenario can be ascribed to the emergence of an imaginary time at $r<2M$. Indeed, a  way to explain the  metric signature change in Eq. \eqref{Lorentzian-Euclidean-Schwarzschild} consists in supposing that the coordinate time $t$  is no longer a real-valued variable inside the black hole. 

This feature can be related to the concept of atemporality, which thus is described as the dynamical mechanism that allows one to avoid the black-hole singularity.

A crucial aspect  consists in proving that the observer in radial free fall takes an infinite amount of  proper time to stop at the event horizon. Following the above equations, $\sigma$ is the proper time in the Lorentzian domain. The behaviour of $\sigma(\eta)$ at $r=2M$ can be thus inferred by letting $\varepsilon$ approach zero. In particular, one finds   $\lim\limits_{\varepsilon \to 0^+} \sigma = + \infty$ and $\lim\limits_{\varepsilon \to 0^-} \sigma =  \infty$, with $\cos (n \, \eta_H)$ finite ($n=3/2,1,1/2$). The first limit means that proper time becomes infinite when $r=2M$, while the second is consistent with the fact that $\sigma$ takes imaginary values for $r<2M$. 
Although it remains true that the (regularized) Kretschmann curvature invariant $\mathcal{K}$  blows up at $r=0$, this investigation shows that this point cannot be reached by infalling radial geodesics. In other words, since $2M \leq r(\sigma) < \infty$,  $\mathcal{K}$ is always bounded along the trajectories followed by radial geodesics, as its maximum value is 
\begin{align}
\label{regularK}
\mathcal{K}(r=2M)=R_{\alpha\beta\mu\nu} R^{\alpha\beta\mu\nu}= \frac{3}{4M^4}\,,
\end{align}
where $R_{\alpha\beta\mu\nu}$ is the Riemann tensor calculated at $r=2M$.  This ties in with a crucial aspects of singularity theorems \cite{Hawking1970} (see also \cite{Magalhaes2024} and references therein): there is no necessary link between curvature invariants that can be measured (and be therefore meaningful for GR) and spacetime singularities. In our case, the geodesic structure is complete. This also holds for the dynamics of radially accelerated particles which share important similarities with the free-falling case. As shown in \cite{Capozziello:2024ucm}, the accelerated observer also halts at the event horizon;  moreover, when $r<2M$, its radial velocity becomes imaginary.
In other words, just like in the case of the geodesic motion, an accelerated body takes an infinite amount of proper time to stop at the event horizon; similarly, a distant observer sees the accelerated particle approaching $r=2M$ in an infinite time.   

\section{Atemporality and Measurement}\label{sec3}

In the signature changing metric, the coordinate time $t$ is no longer a real-valued variable inside the black hole. The singularity is avoided because a massive particle cannot cross the event horizon and would take an infinite amount of time to reach it.\footnote{It is worth noticing that the very same notion of singularity and the debates surrounding its existence is of no use within this new scenario.} In this picture, atemporality is the natural mechanism explaining the transition from a real time dominated region to another one where imaginary time is admitted. 
In this particular context, atemporality can be defined as follows \cite{Capozziello:2024ucm}:

\bigskip
\emph{Atemporality is the dynamical mechanism by which an observer pointing towards the event horizon cannot reach the singularity in $r=0$, because real-valued geodesics and accelerated orbits cannot be prolonged up to there. As a consequence, both the time variable and radial velocity become imaginary inside the black hole. The parameter `measuring' the degree of atemporality is the Kretschmann scalar $\mathcal{K}$, which is related to the mass of the black hole.}
\bigskip

From a conceptual standpoint, it is worth noticing that atemporality is by no means associated to concepts, such as eternity, negation of time and the like. Atemporality is here defined in terms of the process avoiding that classical infalling bodies assume imaginary radial velocity. It is defined as a dynamical mechanism because it is a process that naturally descends from the type of complete geodesics that respond to conservation laws. In other words, any massive object, falling into a black hole, never crosses the event horizon: it is scattered or orbits around it. Beyond the event horizon, time becomes imaginary and then causality cannot be defined. In this sense, atemporality is the mechanism that preserves causality in our world because time, to be physically defined, must be real.

One might ask, at this point, how to reconcile the dynamics described in this paper with the view according to which one just avoids the coordinate singularity by suppressing coordinate time, but still the proper time of the infalling body can be taken into account once it passes the event horizon and falls towards the physical singularity. This view is at odd with the Lorentzian-Euclidean spacetime. However, what is interesting is why it is ``wrong'' from our perspective. The radial velocity of the infalling body becomes imaginary beyond the event horizon, the proper time of a massive body with radial velocity equal to 0 is nothing from the standpoint of a theory of measurement, such as the theory of general relativity, because real-valued geodesics cannot be prolonged over the event horizon and thus become ``incommensurable''. In other words, the fundamental ontology of both GR and SR is conserved in our model, as clocks measuring real time are useless in the Euclidean region. This means that the Lorentzian-Euclidean spacetime shows that where GR is, no singularity arises, with all the consequences that it bears for the understanding of relativistic physics and for Quantum Gravity approaches. The singularity is avoided by a physical process involving the gravitational field and the dynamics of in-falling particles. Atemporality is thus intrinsically related to the emergence of imaginary time, thereby pointing to the non-physical nature of singularities. In other words, atemporality is the signature of a mechanism that generates a forbidden inaccessible region on the ground of a well-defined physical mechanism without arbitrary statements or the appeal to {\it ad hoc} solutions.\footnote{It must be noticed that a complete treatment of atemporality in black holes should also take into account quantum effects. This will be the subject of forthcoming studies.} 
However, in the present paper, we want to suggest that atemporality as presented in \cite{Capozziello:2024ucm} can be related to conservation laws and to the possibility of achieving important results in building up the foundations of a {\it Singularity--Free Physics}.

\section{Atemporality  from Conservation Laws}\label{sec4}

The example discussed in Section 2 can be considered as a particular case in a more general context where atemporality descends from fundamental laws of physics. In the Lorentzian-Euclidean black hole, atemporality is linked to the structure parameters of the black hole, such as its size and mass, which define the event horizon. As in the case of Heisenberg's Uncertainty Principle, atemporality is the signature of a limiting rule for measurements and points to a dynamical mechanism that prevents the loss of causality in the Lorentzian region. In fact, causality is lost when time becomes imaginary, as it happens when a massive particle enters the black hole, and no measurement can be performed inside a black hole.

However, the Schwarzschild radius (related to the mass of all self-gravitating bodies) is a conserved quantity \cite{capozziello2007spherically} and any theory of gravity presents characteristic lengths related to the existence of Noether symmetries \cite{Bernal:2011qz, Bajardi:2022ypn, Capozziello:2017rvz}.
Thus, from the present perspective, atemporality can be identified with the mechanism by means of which a conservation law is {\it always} conserved. Indeed, the avoidance of the singularity can be measured by the Kretschmann scalar that is related to the black hole size, i.e., the larger the black hole, the smaller the value of $\mathcal{K}$ at $r = 2M$ (and vice versa), and offers the possibility of ``measuring'' the degree of atemporality of the system. 

This prediction must be verified, and future observational campaigns can shed new light on how atemporality can be measured. However, from the conceptual standpoint, one can already conclude that atemporality works as a natural mechanism that is implemented in black holes by fixing a boundary condition for a conservation law which is described by the Kretschmann scalar, i.e., the curvature, and the Schwarzschild radius. In other words, atemporality corresponds to the process according to which conservation laws are not violated in the Lorentzian region and at the event horizon, thereby ensuring the possibility of performing physically consistent measurements and of conserving causal connection. 
This amounts to spell out in which sense atemporality is a ``natural'' mechanism avoiding the singularity. When considering in-falling particles freely moving in the radial direction, since one defined the energy as $E=-\varepsilon g_{\mu \nu} \xi^\mu u^\nu$, being $\xi^\mu$ a Killing vector, one is selecting a Noether symmetry. Consequently, the conservation of energy related to the time translation, is preserved up to the event horizon. The velocity of the infalling particle $\dot{r}$ vanishes at the event horizon and becomes imaginary after having crossed it, because the coordinate time $t$ is no longer a real-valued variable inside the black hole. This means that the time translation symmetry and the conservation of energy can hold if and only if the singularity at $r=0$ can be evaded. In fact, the observer never reaches $r=0$, due to the emergence of an imaginary time: as soon as $r<2M$, a change of signature results and the Lorentzian region becomes Euclidean. 
In other words, atemporality stems for a mechanism according to which a conservation law is absolutely preserved otherwise the system loses its physical meaning. Thus, atemporality amounts to preserving causality and to the possibility of performing measurements that are relativistically consistent. 

With the above considerations in mind, we can state the following:\\ 

\textbf{Atemporality Theorem:} {\it Atemporality is the dynamical mechanism which, by preserving a conservation law, allows events in a Lorentzian-Euclidean spacetime to remain causally connected. As a consequence, any singularity is avoided and time can only be defined through real values. If the conservation law is violated, a singularity emerges, time becomes imaginary, and relativistic measurements are impossible.}\\

\textbf{Proof}\\
In the Lorentzian-Euclidean Schwarzschild black hole, any proper observer takes an infinite time to reach the event horizon. The geodesic structure results as complete and this implies that energy and momenta are conserved according to Noether's theorems. This also means that any particle remains in the region $r>2M$.  As a consequence, the Kretschmann curvature invariant is finite and cannot become infinite because observers in free fall and observers with any acceleration approaching the event horizon cannot reach the singularity at $r=0$. In other words, Noether symmetries are preserved for time $t\in \mathcal{R}$ and violated for $t\in \mathcal{I}$. In this context, atemporality means that time never becomes imaginary and singularity is avoided.
$\Box$\\

This theorem portrays atemporality as playing a fundamental role in physics, if one considers that Noether symmetries are always related to the existence of conservation laws. As discussed in \cite{Bajardi:2022ypn}, indeed, physical systems can be consistently defined if a set of conserved quantities is identified. It is worth noticing that relations between conservation laws and Noether symmetries have been widely investigated in the philosophy of physics  \cite{lange2007laws,brading2002symmetry,brading2003symmetries,earman2004laws,Read_Teh_2022}, but we are now explicitly relating them to specific solutions of Einstein equations.

\section{A New Perspective on Singularity--Free Physics?}\label{sec5}

We are now in a position to scrutinize the conceptual implications of the previous considerations. First, it seems possible to identify a natural mechanism dubbed ``atemporality'' applicable to black holes. The remarkable result is that this mechanism is fully consistent with General Relativity and with conservation laws of physics. This aspect has important implications on the study of causality (causal interaction) and the causal structure in the philosophy of spacetime because our model implies that black holes implement the atemporality mechanism to conserve the causal structure of spacetime, i.e., events at different space-time points can be connected by causal curves and causally interact. According to \cite{malament1975does}, ``if the points are connected by a causal curve then it is possible that some signal travels from one to another (along the curve) while at all times moving less than or equal to the maximal possible velocity''. However, as we have seen, the constraint of energy and momentum dictated by Noether’s theorems makes atemporality unavoidable in order to preserve conservation laws: there cannot be causal interaction between an event at a space-time point and the singularity. This account is not only compatible, but verifies what has been suggested in \cite{doi:10.1086/727030} and in \cite{malament1975does}, pp. 111 ff.. In particular, it is clear that causation not only can, but {\it must} be defended in general relativity.

Furthermore, the definition of atemporality as a mechanism to avoid singularities rules out the long-standing misconception according to which atemporality equates a static metaphysical notion of eternity. In the last twenty centuries, the notion of atemporality has designated this concept as the negation of time, being the latter understood in terms of an absolute container, an {\it a priori} subjective form of intuition or as illusion. Thanks to the present approach, we have seen that modeling the physical world admits both temporal and atemporal forms of representing motion, i. e., becoming. This result impacts, of course, the philosophy of cosmology, because it hints to the fact that our universe can be modeled as a system embedding evolving phenomena, as well as transitions governed by atemporality, i. e., connected topology or signature change descend from the preservation of conservation laws.

Second, debates in the philosophy of physics \cite{Read_Teh_2022} have already underscored that Noether theorems acquire a fundamental meaning for the analysis of the symmetry groups and the invariants of equations in general relativity \cite{Haro_2022}, as well as for establishing the status of conservation laws and gauge theory.\footnote{Brown, in \cite{Brown_2022}, suggested that the explanatory implication from symmetries to conservation laws cannot be extrapolated from the Noether first theorem, even if symmetries have pragmatic priority assigned and play a fundamental heuristic role in physics.} In this sense, they can be read as necessary principles of physics, but, as the so-called ``Noether Symmetry Approach'' \cite{Bajardi:2022ypn, Capozziello:1996bi} has shown, they also provide the heuristic framework to find exact solutions of the field equations and to implement Extended Theories of Gravity \cite{capozziello2007spherically, Faraoni:2010pgm}. Furthermore, as Gomes, Roberts, and Butterfield \cite{Gomes_Roberts_Butterfield_2022} have convincingly shown, Noether theorems represent the rationale for gauge symmetry to provide a path to build appropriate dynamical theories. In this sense, it would be interesting to extend in the future our analysis of atemporality towards these areas of philosophical investigation. For the time being and within the scope of the present work, we expounded in which sense we can talk of a derivation of atemporality from conservation laws. The Lorentzian-Euclidean black hole is a model that not only allows one to explore the hidden implications of GR, but it also reinterprets the notion of imaginary time by means of the description of a new dynamics at the event horizon.  
In this respect, future research programs should mark a further step with respect to \cite{Baez_2022} in clarifying the link between general relativity, quantum and statistical mechanics through a dynamical correspondence. These issues will be discussed in a separate work.

Third, an important part of \cite{Capozziello:2024ucm} impacts the field of philosophy of time because it forces philosophers and physicists to think of imaginary time and atemporality in new terms. On the one hand, imaginary time should be treated on the same footing as real time, but it clearly assumes a different meaning. It plays the same role that real time has in the Lorentzian region, but it does so in the Euclidean one. Furthermore, imaginary time cannot equate atemporality and its structure differs from that of real time. Questions concerning structure, derivation and ``shape'' of imaginary time can be open, together with the more fundamental question of what imaginary time could ``measure'' in a Euclidean region and which ontological status it possesses.
On the other hand, if atemporality is defined in terms of a process or dynamical mechanism, it also opens the path for the quest of a suitable taxonomy that accounts for its ontological and/or epistemological status. Put it in simple terms, it could seem a {\it contradictio in adiecto} to talk about a dynamical mechanism in atemporal terms. However, as shown in \cite{de2023re}, it is perfectly possible to think of specific forms of discrete atemporal ordering, e.g., instantaneity, without contradiction, when one has to talk about phase transitions such as geometrogenesis in Quantum Gravity scenarios. Thanks to the present work, one can now add another piece to this taxonomy and talk about atemporal ordering according to principles of continuous action to be explored in extreme gravitational conditions in agreement with conservation laws.

\section{Conclusions}\label{sec6}
According to James Read, there are still several open questions in Black Hole physics that must be addressed, e.g., the study of their thermodynamical nature, a clear understanding of the conditions under which material fields come to be adapted to a given spacetime geometry and the understanding of the physics and metaphysics of singularities in view of singularity resolution in Quantum Gravity or to clarify the black hole information loss paradox (for a longer list, see \cite{pittphilsci24520}). The fact that singularities can be avoided from the general relativistic standpoint in Lorentzian-Euclidean black holes by means of atemporality can enrich the list by further stimulating the application of this concept beyond the case of in-falling massive bodies and stationary black holes. This dynamical mechanism, dubbed ``atemporality'', prevents both the emergence of black hole singularities and the violation of conservation laws. 
Essentially, this is due to the fact that the mechanism itself derives from conservation laws of physics and consists in preserving the physical meaning of the system only in the region where time is defined as real. By crossing the event horizon, time becomes imaginary and causality is lost. 
As pointed out, this is only a particular case of a more general framework where conservation laws allow to define a {\it Singularity--Free Physics}: it means that as soon as atemporality prevents that a physical system assumes imaginary time, any singularity is removed and causality is preserved. In other words, at least at the classical level, it is impossible that a physical system is not measurable and loses its predictability (causality in the classical sense of succession of real instants). Thus, the notion of atemporality is restoring Einstein's idea that physics responds to a causal structure of spacetime according to which there is no room for singularities.

\backmatter

\subsection*{Acknowledgements}
SDB acknowledges the funding and support from the European Union's Horizon 2020 research and innovation program under Grant Agreement No. 758145 -- PROTEUS and the project (2021-0567-COSMOS) funded by Cariplo Foundation. SC acknowledges the support of Istituto Nazionale di Fisica Nucleare (INFN) Sez. di Napoli, Iniziative Specifiche QGSKY.  EB acknowledges the support of Istituto Nazionale di Fisica Nucleare (INFN), {\it Iniziative Specifiche} MOONLIGHT2. 
This paper is based upon work from COST Action CA21136 {\it Addressing observational tensions in cosmology with systematics and fundamental physics} (CosmoVerse) supported by COST (European Cooperation in Science and Technology).

\bibliography{sn-bibliography}


\begin{thebibliography}{33}
\ifx \bisbn   \undefined \def \bisbn  #1{ISBN #1}\fi
\ifx \binits  \undefined \def \binits#1{#1}\fi
\ifx \bauthor  \undefined \def \bauthor#1{#1}\fi
\ifx \batitle  \undefined \def \batitle#1{#1}\fi
\ifx \bjtitle  \undefined \def \bjtitle#1{#1}\fi
\ifx \bvolume  \undefined \def \bvolume#1{\textbf{#1}}\fi
\ifx \byear  \undefined \def \byear#1{#1}\fi
\ifx \bissue  \undefined \def \bissue#1{#1}\fi
\ifx \bfpage  \undefined \def \bfpage#1{#1}\fi
\ifx \blpage  \undefined \def \blpage #1{#1}\fi
\ifx \burl  \undefined \def \burl#1{\textsf{#1}}\fi
\ifx \doiurl  \undefined \def \doiurl#1{\url{https://doi.org/#1}}\fi
\ifx \betal  \undefined \def \betal{\textit{et al.}}\fi
\ifx \binstitute  \undefined \def \binstitute#1{#1}\fi
\ifx \binstitutionaled  \undefined \def \binstitutionaled#1{#1}\fi
\ifx \bctitle  \undefined \def \bctitle#1{#1}\fi
\ifx \beditor  \undefined \def \beditor#1{#1}\fi
\ifx \bpublisher  \undefined \def \bpublisher#1{#1}\fi
\ifx \bbtitle  \undefined \def \bbtitle#1{#1}\fi
\ifx \bedition  \undefined \def \bedition#1{#1}\fi
\ifx \bseriesno  \undefined \def \bseriesno#1{#1}\fi
\ifx \blocation  \undefined \def \blocation#1{#1}\fi
\ifx \bsertitle  \undefined \def \bsertitle#1{#1}\fi
\ifx \bsnm \undefined \def \bsnm#1{#1}\fi
\ifx \bsuffix \undefined \def \bsuffix#1{#1}\fi
\ifx \bparticle \undefined \def \bparticle#1{#1}\fi
\ifx \barticle \undefined \def \barticle#1{#1}\fi
\bibcommenthead
\ifx \bconfdate \undefined \def \bconfdate #1{#1}\fi
\ifx \botherref \undefined \def \botherref #1{#1}\fi
\ifx \url \undefined \def \url#1{\textsf{#1}}\fi
\ifx \bchapter \undefined \def \bchapter#1{#1}\fi
\ifx \bbook \undefined \def \bbook#1{#1}\fi
\ifx \bcomment \undefined \def \bcomment#1{#1}\fi
\ifx \oauthor \undefined \def \oauthor#1{#1}\fi
\ifx \citeauthoryear \undefined \def \citeauthoryear#1{#1}\fi
\ifx \endbibitem  \undefined \def \endbibitem {}\fi
\ifx \bconflocation  \undefined \def \bconflocation#1{#1}\fi
\ifx \arxivurl  \undefined \def \arxivurl#1{\textsf{#1}}\fi
\csname PreBibitemsHook\endcsname

\bibitem[\protect\citeauthoryear{Earman}{1995}]{Earman1995-EARBCW-3}
\begin{bbook}
\bauthor{\bsnm{Earman}, \binits{J.}}:
\bbtitle{Bangs, Crunches, Whimpers, and Shrieks: Singularities and Acausalities
  in Relativistic Spacetimes}.
\bpublisher{Oxford University Press},
\blocation{Oxford}
(\byear{1995})
\end{bbook}
\endbibitem

\bibitem[\protect\citeauthoryear{Curiel}{2019}]{curiel2019many}
\begin{barticle}
\bauthor{\bsnm{Curiel}, \binits{E.}}:
\batitle{The many definitions of a black hole}.
\bjtitle{Nature Astronomy}
\bvolume{3}(\bissue{1}),
\bfpage{27}--\blpage{34}
(\byear{2019})
\end{barticle}
\endbibitem

\bibitem[\protect\citeauthoryear{Ellis et~al.}{1992}]{ellis1992change}
\begin{barticle}
\bauthor{\bsnm{Ellis}, \binits{G.}},
\bauthor{\bsnm{Sumeruk}, \binits{A.}},
\bauthor{\bsnm{Coule}, \binits{D.}},
\bauthor{\bsnm{Hellaby}, \binits{C.}}:
\batitle{Change of signature in classical relativity}.
\bjtitle{Classical and Quantum Gravity}
\bvolume{9}(\bissue{6}),
\bfpage{1535}
(\byear{1992})
\end{barticle}
\endbibitem

\bibitem[\protect\citeauthoryear{Ellis}{1992}]{ellis1992covariant}
\begin{barticle}
\bauthor{\bsnm{Ellis}, \binits{G.F.}}:
\batitle{Covariant change of signature in classical relativity}.
\bjtitle{General relativity and gravitation}
\bvolume{24},
\bfpage{1047}--\blpage{1068}
(\byear{1992})
\end{barticle}
\endbibitem

\bibitem[\protect\citeauthoryear{Li and Singh}{2023}]{li2023loop}
\begin{bchapter}
\bauthor{\bsnm{Li}, \binits{B.-F.}},
\bauthor{\bsnm{Singh}, \binits{P.}}:
\bctitle{Loop quantum cosmology: Physics of singularity resolution and its
  implications}.
In: \bbtitle{Handbook of Quantum Gravity},
pp. \bfpage{1}--\blpage{55}.
\bpublisher{Springer},
\blocation{Singapore}
(\byear{2023})
\end{bchapter}
\endbibitem

\bibitem[\protect\citeauthoryear{Vidotto}{2022}]{vidotto2022time}
\begin{bchapter}
\bauthor{\bsnm{Vidotto}, \binits{F.}}:
\bctitle{Time, space and matter in the primordial universe}.
In: \bbtitle{Advances in Cosmology: Science-Art-Philosophy},
pp. \bfpage{333}--\blpage{344}.
\bpublisher{Springer},
\blocation{Cham}
(\byear{2022})
\end{bchapter}
\endbibitem

\bibitem[\protect\citeauthoryear{Oriti}{2023}]{oriti2023complex}
\begin{bchapter}
\bauthor{\bsnm{Oriti}, \binits{D.}}:
\bctitle{The complex timeless emergence of time in quantum gravity}.
In: \bbtitle{TIME AND SCIENCE: Volume 3: Physical Sciences and Cosmology},
pp. \bfpage{137}--\blpage{175}.
\bpublisher{World Scientific},
\blocation{Europe}
(\byear{2023})
\end{bchapter}
\endbibitem

\bibitem[\protect\citeauthoryear{Gielen and
  Santacruz}{2023}]{gielen2023stationary}
\begin{barticle}
\bauthor{\bsnm{Gielen}, \binits{S.}},
\bauthor{\bsnm{Santacruz}, \binits{R.}}:
\batitle{Stationary cosmology in group field theory}.
\bjtitle{Physical Review D}
\bvolume{108}(\bissue{2}),
\bfpage{026001}
(\byear{2023})
\end{barticle}
\endbibitem

\bibitem[\protect\citeauthoryear{Zee}{2010}]{Zee:2010qce}
\begin{bbook}
\bauthor{\bsnm{Zee}, \binits{A.}}:
\bbtitle{{Quantum Field Theory in a Nutshell: Second Edition}}.
\bpublisher{Princeton University Press},
\blocation{Princeton}
(\byear{2010})
\end{bbook}
\endbibitem

\bibitem[\protect\citeauthoryear{Hartle et~al.}{2008}]{Hartle:2008ng}
\begin{barticle}
\bauthor{\bsnm{Hartle}, \binits{J.B.}},
\bauthor{\bsnm{Hawking}, \binits{S.W.}},
\bauthor{\bsnm{Hertog}, \binits{T.}}:
\batitle{{The Classical Universes of the No-Boundary Quantum State}}.
\bjtitle{Phys. Rev. D}
\bvolume{77},
\bfpage{123537}
(\byear{2008})
\doiurl{10.1103/PhysRevD.77.123537}
{\href{https://arxiv.org/abs/0803.1663}{{arXiv:0803.1663}}}
{[hep-th]}
\end{barticle}
\endbibitem

\bibitem[\protect\citeauthoryear{Capozziello
  et~al.}{2024}]{Capozziello:2024ucm}
\begin{barticle}
\bauthor{\bsnm{Capozziello}, \binits{S.}},
\bauthor{\bsnm{De~Bianchi}, \binits{S.}},
\bauthor{\bsnm{Battista}, \binits{E.}}:
\batitle{{Avoiding singularities in Lorentzian-Euclidean black holes: The role
  of~atemporality}}.
\bjtitle{Phys. Rev. D}
\bvolume{109}(\bissue{10}),
\bfpage{104060}
(\byear{2024})
\doiurl{10.1103/PhysRevD.109.104060}
{\href{https://arxiv.org/abs/2404.17267}{{arXiv:2404.17267}}}
{[gr-qc]}
\end{barticle}
\endbibitem

\bibitem[\protect\citeauthoryear{Chandrasekhar}{1983}]{Chandrasekhar1985}
\begin{bbook}
\bauthor{\bsnm{Chandrasekhar}, \binits{S.}}:
\bbtitle{{The Mathematical Theory of Black Holes}}.
\bpublisher{Oxford University Press},
\blocation{Oxford}
(\byear{1983})
\end{bbook}
\endbibitem

\bibitem[\protect\citeauthoryear{{Hawking} and {Penrose}}{1970}]{Hawking1970}
\begin{barticle}
\bauthor{\bsnm{{Hawking}}, \binits{S.W.}},
\bauthor{\bsnm{{Penrose}}, \binits{R.}}:
\batitle{{The Singularities of Gravitational Collapse and Cosmology}}.
\bjtitle{Proceedings of the Royal Society of London Series A}
\bvolume{314}(\bissue{1519}),
\bfpage{529}--\blpage{548}
(\byear{1970})
\doiurl{10.1098/rspa.1970.0021}
\end{barticle}
\endbibitem

\bibitem[\protect\citeauthoryear{Magalh\~aes et~al.}{2024}]{Magalhaes2024}
\begin{botherref}
\oauthor{\bsnm{Magalh\~aes}, \binits{R.B.}},
\oauthor{\bsnm{Ribeiro}, \binits{G.P.}},
\oauthor{\bsnm{Lima~Junior}, \binits{H.C.D.}},
\oauthor{\bsnm{Olmo}, \binits{G.J.}},
\oauthor{\bsnm{Crispino}, \binits{L.C.B.}}:
{Singular space-times with bounded algebraic curvature scalars}
(2024)
{\href{https://arxiv.org/abs/2401.12779}{{arXiv:2401.12779}}}
{[gr-qc]}
\end{botherref}
\endbibitem

\bibitem[\protect\citeauthoryear{Capozziello
  et~al.}{2007}]{capozziello2007spherically}
\begin{barticle}
\bauthor{\bsnm{Capozziello}, \binits{S.}},
\bauthor{\bsnm{Stabile}, \binits{A.}},
\bauthor{\bsnm{Troisi}, \binits{A.}}:
\batitle{Spherically symmetric solutions in $f(r)$ gravity via the noether
  symmetry approach}.
\bjtitle{Classical and Quantum Gravity}
\bvolume{24}(\bissue{8}),
\bfpage{2153}
(\byear{2007})
\end{barticle}
\endbibitem

\bibitem[\protect\citeauthoryear{Bernal et~al.}{2011}]{Bernal:2011qz}
\begin{barticle}
\bauthor{\bsnm{Bernal}, \binits{T.}},
\bauthor{\bsnm{Capozziello}, \binits{S.}},
\bauthor{\bsnm{Hidalgo}, \binits{J.C.}},
\bauthor{\bsnm{Mendoza}, \binits{S.}}:
\batitle{{Recovering MOND from extended metric theories of gravity}}.
\bjtitle{Eur. Phys. J. C}
\bvolume{71},
\bfpage{1794}
(\byear{2011})
\doiurl{10.1140/epjc/s10052-011-1794-z}
{\href{https://arxiv.org/abs/1108.5588}{{arXiv:1108.5588}}}
{[astro-ph.CO]}
\end{barticle}
\endbibitem

\bibitem[\protect\citeauthoryear{Bajardi and
  Capozziello}{2022}]{Bajardi:2022ypn}
\begin{bbook}
\bauthor{\bsnm{Bajardi}, \binits{F.}},
\bauthor{\bsnm{Capozziello}, \binits{S.}}:
\bbtitle{{Noether Symmetries in Theories of Gravity}}.
\bsertitle{Cambridge Monographs on Mathematical Physics}.
\bpublisher{Cambridge University Press},
\blocation{Cambridge}
(\byear{2022}).
\doiurl{10.1017/9781009208727}
\end{bbook}
\endbibitem

\bibitem[\protect\citeauthoryear{Capozziello
  et~al.}{2017}]{Capozziello:2017rvz}
\begin{barticle}
\bauthor{\bsnm{Capozziello}, \binits{S.}},
\bauthor{\bsnm{Jovanovi\'c}, \binits{P.}},
\bauthor{\bsnm{Jovanovi\'c}, \binits{V.B.}},
\bauthor{\bsnm{Borka}, \binits{D.}}:
\batitle{{Addressing the missing matter problem in galaxies through a new
  fundamental gravitational radius}}.
\bjtitle{JCAP}
\bvolume{06},
\bfpage{044}
(\byear{2017})
\doiurl{10.1088/1475-7516/2017/06/044}
{\href{https://arxiv.org/abs/1702.03430}{{arXiv:1702.03430}}}
{[gr-qc]}
\end{barticle}
\endbibitem

\bibitem[\protect\citeauthoryear{Lange}{2007}]{lange2007laws}
\begin{barticle}
\bauthor{\bsnm{Lange}, \binits{M.}}:
\batitle{Laws and meta-laws of nature: Conservation laws and symmetries}.
\bjtitle{Studies in History and Philosophy of Science Part B: Studies in
  History and Philosophy of Modern Physics}
\bvolume{38}(\bissue{3}),
\bfpage{457}--\blpage{481}
(\byear{2007})
\end{barticle}
\endbibitem

\bibitem[\protect\citeauthoryear{Brading}{2002}]{brading2002symmetry}
\begin{barticle}
\bauthor{\bsnm{Brading}, \binits{K.A.}}:
\batitle{Which symmetry? noether, weyl, and conservation of electric charge}.
\bjtitle{Studies in History and Philosophy of Science Part B: Studies in
  History and Philosophy of Modern Physics}
\bvolume{33}(\bissue{1}),
\bfpage{3}--\blpage{22}
(\byear{2002})
\end{barticle}
\endbibitem

\bibitem[\protect\citeauthoryear{Brading and
  Brown}{2003}]{brading2003symmetries}
\begin{botherref}
\oauthor{\bsnm{Brading}, \binits{K.}},
\oauthor{\bsnm{Brown}, \binits{H.R.}}:
Symmetries and noether’s theorems.
Symmetries in physics: Philosophical reflections,
89--109
(2003)
\end{botherref}
\endbibitem

\bibitem[\protect\citeauthoryear{Earman}{2004}]{earman2004laws}
\begin{barticle}
\bauthor{\bsnm{Earman}, \binits{J.}}:
\batitle{Laws, symmetry, and symmetry breaking: Invariance, conservation
  principles, and objectivity}.
\bjtitle{Philosophy of Science}
\bvolume{71}(\bissue{5}),
\bfpage{1227}--\blpage{1241}
(\byear{2004})
\end{barticle}
\endbibitem

\bibitem[\protect\citeauthoryear{Read and Teh}{2022}]{Read_Teh_2022}
\begin{bbook}
\bauthor{\bsnm{Read}, \binits{J.}},
\bauthor{\bsnm{Teh}, \binits{N.J.}}:
\bbtitle{The Philosophy and Physics of Noether's Theorems: A Centenary Volume}.
\bpublisher{Cambridge University Press},
\blocation{Cambridge}
(\byear{2022})
\end{bbook}
\endbibitem

\bibitem[\protect\citeauthoryear{Malament}{1975}]{malament1975does}
\begin{botherref}
\oauthor{\bsnm{Malament}, \binits{D.B.}}:
Does the causal structure of space-time determine its geometry
(1975)
\end{botherref}
\endbibitem

\bibitem[\protect\citeauthoryear{Ram\'{\i}rez et~al.}{}]{doi:10.1086/727030}
\begin{botherref}
\oauthor{\bsnm{Ram\'{\i}rez}, \binits{S.M.}},
\oauthor{\bsnm{Read}, \binits{J.A.M.}},
\oauthor{\bsnm{Paez}, \binits{A.}}:
Causation and the conservation of energy in general relativity.
The British Journal for the Philosophy of Science
{\href{https://arxiv.org/abs/https://doi.org/10.1086/727030}{{https://doi.org/10.1086/727030}}}
\end{botherref}
\endbibitem

\bibitem[\protect\citeauthoryear{De~Haro}{2022}]{Haro_2022}
\begin{bbook}
\bauthor{\bsnm{De~Haro}, \binits{S.}}:
In: \beditor{\bsnm{Read}, \binits{J.}},
\beditor{\bsnm{Teh}, \binits{N.J.E.}} (eds.)
\bbtitle{Noether's Theorems and Energy in General Relativity},
pp. \bfpage{197}--\blpage{256}.
\bpublisher{Cambridge University Press},
\blocation{Cambridge}
(\byear{2022})
\end{bbook}
\endbibitem

\bibitem[\protect\citeauthoryear{Brown}{2022}]{Brown_2022}
\begin{bbook}
\bauthor{\bsnm{Brown}, \binits{H.R.}}:
In: \beditor{\bsnm{Read}, \binits{J.}},
\beditor{\bsnm{Teh}, \binits{N.J.E.}} (eds.)
\bbtitle{Do Symmetries `Explain' Conservation Laws? The Modern Converse Noether
  Theorem vs Pragmatism},
pp. \bfpage{144}--\blpage{168}.
\bpublisher{Cambridge University Press},
\blocation{Cambridge}
(\byear{2022})
\end{bbook}
\endbibitem

\bibitem[\protect\citeauthoryear{Capozziello et~al.}{1996}]{Capozziello:1996bi}
\begin{barticle}
\bauthor{\bsnm{Capozziello}, \binits{S.}},
\bauthor{\bsnm{De~Ritis}, \binits{R.}},
\bauthor{\bsnm{Rubano}, \binits{C.}},
\bauthor{\bsnm{Scudellaro}, \binits{P.}}:
\batitle{{Noether symmetries in cosmology}}.
\bjtitle{Riv. Nuovo Cim.}
\bvolume{19N4},
\bfpage{1}--\blpage{114}
(\byear{1996})
\doiurl{10.1007/BF02742992}
\end{barticle}
\endbibitem

\bibitem[\protect\citeauthoryear{Capozziello and
  Faraoni}{2011}]{Faraoni:2010pgm}
\begin{bbook}
\bauthor{\bsnm{Capozziello}, \binits{S.}},
\bauthor{\bsnm{Faraoni}, \binits{V.}}:
\bbtitle{{Beyond Einstein Gravity}: {A Survey of Gravitational Theories for
  Cosmology and Astrophysics}}.
\bpublisher{Springer},
\blocation{Dordrecht}
(\byear{2011}).
\doiurl{10.1007/978-94-007-0165-6}
\end{bbook}
\endbibitem

\bibitem[\protect\citeauthoryear{Gomes
  et~al.}{2022}]{Gomes_Roberts_Butterfield_2022}
\begin{bbook}
\bauthor{\bsnm{Gomes}, \binits{H.}},
\bauthor{\bsnm{Roberts}, \binits{B.W.}},
\bauthor{\bsnm{Butterfield}, \binits{J.}}:
In: \beditor{\bsnm{Read}, \binits{J.}},
\beditor{\bsnm{Teh}, \binits{N.J.E.}} (eds.)
\bbtitle{The Gauge Argument: A Noether Reason},
pp. \bfpage{354}--\blpage{376}.
\bpublisher{Cambridge University Press},
\blocation{Cambridge}
(\byear{2022})
\end{bbook}
\endbibitem

\bibitem[\protect\citeauthoryear{Baez}{2022}]{Baez_2022}
\begin{bbook}
\bauthor{\bsnm{Baez}, \binits{J.C.}}:
In: \beditor{\bsnm{Read}, \binits{J.}},
\beditor{\bsnm{Teh}, \binits{N.J.E.}} (eds.)
\bbtitle{Getting to the Bottom of Noether’s Theorem},
pp. \bfpage{66}--\blpage{99}.
\bpublisher{Cambridge University Press},
\blocation{Cambridge}
(\byear{2022})
\end{bbook}
\endbibitem

\bibitem[\protect\citeauthoryear{De~Bianchi and Gabbanelli}{2023}]{de2023re}
\begin{botherref}
\oauthor{\bsnm{De~Bianchi}, \binits{S.}},
\oauthor{\bsnm{Gabbanelli}, \binits{L.}}:
Re-thinking geometrogenesis: instantaneity in quantum gravity scenarios.
J. Phys.: Conf. Ser.
\textbf{2533}
(2023)
\end{botherref}
\endbibitem

\bibitem[\protect\citeauthoryear{Read}{2025}]{pittphilsci24520}
\begin{botherref}
\oauthor{\bsnm{Read}, \binits{J.}}:
23 open problems in the philosophy of physics
(2025).
\url{https://philsci-archive.pitt.edu/24520/}
\end{botherref}
\endbibitem

\end{thebibliography}

\end{document}